\documentclass[prd,twocolumn,superscriptaddress,nofootinbib,floatfix,preprintnumbers]{revtex4}
\pdfoutput=1

\usepackage{amssymb, bm, amsmath, graphicx, physymb}
\usepackage[colorlinks=true,linktocpage=true,linkcolor=blue,citecolor=blue]{hyperref}
\usepackage[usenames,dvipsnames]{color}
\usepackage{slashed}
\usepackage[utf8]{inputenc}
\usepackage[T1]{fontenc}
\usepackage{enumerate}
\usepackage{amsfonts}
\usepackage{mathtools}
\usepackage{latexsym}
\usepackage{hyperref}
\usepackage{epstopdf} 
\usepackage{bbm}
\usepackage{soul}
\usepackage[normalem]{ulem}
\usepackage{braket}
\usepackage{nccmath}
\usepackage{graphicx}

\usepackage{xcolor}

\usepackage[e]{esvect}

\newcommand{\secn}[1]{Section~1}
\newcommand{\appn}[1]{Appendix~1}

\long\def\comment#1{ }

\def\and{\quad\text{and}\quad}

\def\0{{\boldsymbol 0}}
\def\1{{\boldsymbol 1}}

\def\b{{\boldsymbol b}}
\def\0{{\boldsymbol 0}}

\renewcommand\a{\alpha}
\renewcommand\b{\beta}
\renewcommand\d{\delta}
\renewcommand\k{\kappa}
\renewcommand\l{\lambda}

\renewcommand\o{\omega}

\newcommand\m{\mu}
\newcommand\n{\nu}

\newcommand\s{\sigma}

\renewcommand{\vec}{\boldsymbol}

\renewcommand{\part}{{\rm part}}

\newcommand{\be}{\begin{equation}}
\newcommand{\ee}{\end{equation}}
\newcommand{\bes}{\begin{subequations}}
\newcommand{\ees}{\end{subequations}}
\newcommand{\bea}{\begin{eqnarray}}
\newcommand{\eea}{\end{eqnarray}}

\newcommand{\pa}{\partial}

\newcommand{\beq}{\begin{equation}}
\newcommand{\eeq}{\end{equation}}

\makeatletter
\newsavebox{\@brx}
\newcommand{\llangle}[1][]{\savebox{\@brx}{\(\m@th{#1\langle}\)}%
  \mathopen{\copy\@brx\mkern2mu\kern-0.9\wd\@brx\usebox{\@brx}}}
\newcommand{\rrangle}[1][]{\savebox{\@brx}{\(\m@th{#1\rangle}\)}%
  \mathclose{\copy\@brx\mkern2mu\kern-0.9\wd\@brx\usebox{\@brx}}}
\makeatother

\begin{document}

\title{Chiral Vortical Instability}

\author{Shuai Wang}
\email[Email: ]{swang19@fudan.edu.cn}
\affiliation{Physics Department and Center for Particle Physics and Field Theory, Fudan University, Shanghai 200438, China}

\author{Koichi Hattori}
\email[Email: ]{koichi.hattori@zju.edu.cn}
\affiliation{Zhejiang Institute of Modern Physics, Department of Physics, Zhejiang University,
Hangzhou, Zhejiang 310027, China}
\affiliation{Research Center for Nuclear Physics, Osaka University, 
10-1 Mihogaoka, Ibaraki, Osaka 567-0047, Japan}

\author{Xu-Guang Huang}
\email[Email: ]{huangxuguang@fudan.edu.cn}
\affiliation{Physics Department and Center for Particle Physics and Field Theory, Fudan University, Shanghai 200438, China}
\affiliation{Key Laboratory of Nuclear Physics and Ion-beam Application (MOE), Fudan University, Shanghai 200433, China}
\affiliation{Shanghai Research Center for Theoretical Nuclear Physics,
National Natural Science Foundation of China and Fudan University, Shanghai 200438, China}

\author{Andrey V. Sadofyev}
\email[Email: ]{sadofyev@lip.pt}
\affiliation{LIP, Av. Prof. Gama Pinto, 2, P-1649-003 Lisboa, Portugal}

\begin{abstract}
We revisit the collective modes of chiral matter described by the second-order chiral hydrodynamics, noticing that chiral shear waves (CSWs) may become unstable for momenta above a characteristic scale. In the absence of sufficient dissipation, this instability emerges within the hydrodynamic regime, depending on the interplay between shear viscosity and the anomalous vortical contribution to the stress-energy tensor at second order in hydrodynamic expansion. We show that this instability generates helical flows and name it the {\it chiral vortical instability} (CVI). Alongside the chiral plasma and magnetovortical instabilities, CVI tends to transfer initial microscopic chirality into macroscopic helicities, which combine into a generalized axial charge. We further find that an elementary static Gromeka–Arnold–Beltrami–Childress flow, corresponding to a CSW at a specific momentum, solves the full nonlinear equations of second-order chiral hydrodynamics, whereas global rotation of a chiral medium is not a solution. This observation supports the relevance of CVI beyond the hydrodynamic regime. Finally, we briefly note that CVI may have multiple phenomenological implications across various systems, including QCD matter produced in heavy-ion collisions and primordial plasma in the early Universe.
\end{abstract}

\maketitle

\section{Introduction}

The dynamics of the axial charge is known to be modified in chiral media where it mixes with the macroscopic helicities of the field and flow configurations \cite{Avdoshkin:2014gpa, Yamamoto:2015gzz, Avkhadiev:2017fxj, Kirilin:2017tdh, Wiegmann:2022syo, Manuel:2022tck, Nastase:2022aps}. In such systems, the axial anomaly is manifested at the macroscopic level through a class of novel transport phenomena, the chiral anomalous effects, for a review see e.g. \cite{Kharzeev:2015znc, Huang:2015oca, Hattori:2023egw}. These effects contribute to the axial current:
\begin{align}
J_5^\a=n_5 u^\a+\xi_B B^\a+\xi_\o \o^\a\,,
\end{align}
where $\o^\a=\frac{1}{2}\epsilon^{\a\b\m\n}u_\b\pa_\m u_\n$ is the flow vorticity, $B^\a=\frac{1}{2}\epsilon^{\a\b\m\n}u_\b F_{\m\n}$ is the magnetic field in the rest frame of the fluid element, and $\xi_B$ and $\xi_\o$ are the corresponding anomalous kinetic coefficients. Thus, the generalized axial charge conservation \cite{Avdoshkin:2014gpa} involves not only the microscopic chiral density but also contributions accounting for the macroscopic topology of the matter fields
\begin{align}
\pa_t\left\{N_5+\mathcal{H}_{fh}+\mathcal{H}_{mfh}+\mathcal{H}_{mh}\right\}=0\,,
\end{align}
where we assume that there is no axial current at infinity. In the limit of non-relativistic flow, $N_5=\int_x n_5$ corresponds to the microscopic part of the axial charge, $\mathcal{H}_{fh}=\int_x \xi_\o\, \vec{v}\cdot\vec{\o}$ is the flow helicity, $\mathcal{H}_{mfh}=\int_x \xi_B\, \vec{v}\cdot\vec{B}$ is the mixed flow-magnetic field helicity, and $\mathcal{H}_{mh}=\frac{C}{2}\int_x \vec{A}\cdot\vec{B}$ is the magnetic field helicity, accounting for the presence of the axial anomaly $\pa_\a J_5^\a=C\vec{E}\cdot\vec{B}$.

The combination of microscopic chirality and macroscopic helicities in a single generalized axial charge suggests the emergence of a family of instabilities, provided that the different contributions can transform into one another. Indeed, in an equilibrium state one may expect the degrees of freedom to contribute to all the possible forms of any conserved charge, and if the evolution starts with, say, only the microscopic chirality, the system will rapidly evolve to redistribute it over all the forms of the axial charge \cite{Avdoshkin:2014gpa}. Thus, the initial dynamics of a chiral system with the axial charge fully stored in one of its forms is expected to be unstable, stabilizing at later times depending on infrared features of the system, such as its size, for a detailed discussion see \cite{Khaidukov:2013sja, Hirono:2015rla, Hirono:2016jps, Kirilin:2017tdh}. 

A renowned example of such an instability, the so-called chiral plasma instability (CPI), corresponds to a fast transfer of the initial chirality to $\mathcal{H}_{mh}$ absent in the initial state of the system. It is widely discussed in the literature, along with related phenomena, and is argued to have multiple theoretical and phenomenological implications \cite{Rubakov:1985nk, Joyce:1997uy,Giovannini:1997eg,Boyarsky:2011uy, Tashiro:2012mf, Akamatsu:2013pjd, Yamamoto:2015gzz, Manuel:2015zpa, Bhatt:2015ewa, Buividovich:2015jfa, Zakharov:2016lhp, Gorbar:2016qfh, Yamamoto:2016xtu, Hattori:2017usa, Tuchin:2018sqe, Masada:2018swb, Abbaslu:2019yiy, Tuchin:2019gkg, Abbaslu:2020xfn, Brandenburg:2023aco,Brandenburg:2023rul}, for a recent review see \cite{Kamada:2022nyt,
Hattori:2023egw}. Note also that the inverse of this process has received comparatively less attention but can serve as a mechanism for chirality generation or baryogenesis from the initial magnetic helicity, for example, through reconnections of magnetic flux tubes \cite{Giovannini:1997eg, Giovannini:1999wv, Giovannini:1999by, Hirono:2016jps, Kirilin:2017tdh}. It has been shown \cite{Wang:2023imu} that the equations of the magnetohydrodynamics also become unstable in the presence of chiral vortical effect (CVE) in the electric current. This chiral magnetovortical instability (CMVI) results in generation of mixed helicity $\mathcal{H}_{mfh}$ from initial chirality of the matter.

However, despite these advances, little is known about the other channels of this unstable behavior of chiral media, and particularly about the one when the microscopic chirality $N_5$ is pumped into the flow helicity $\mathcal{H}_{fh}$, see \cite{Avdoshkin:2014gpa, Yamamoto:2015gzz, Kirilin:2017tdh} for earlier discussions. 
In this work, we continue studying the chiral instabilities and derive the explicit form of the vortical channel, which we title the {\it chiral vortical instability} (CVI). We further identify the spectrum of the corresponding unstable modes, the chiral shear waves (CSWs) \cite{Sahoo:2009yq}, and show that they take the form of Beltrami flows with exponentially growing amplitude for sufficiently large momenta. Finally, we comment on the evolution of the CVI, its sensitivity to the system properties, compare its features with CPI and CMVI, and briefly discuss potential phenomenological implications of the CVI. Throughout this article, we use the most positive signature for the metric, $g_{\mu\nu}={\rm diag}(-1, 1, 1, 1)$.

\section{The second order Chiral hydrodynamics and its linearized modes}

Starting with the linearized second order chiral hydrodynamics in the Landau frame and closely following its holographic derivation in \cite{Erdmenger:2008rm, Banerjee:2008th}, we consider the spectrum of the matter perturbations in a single chirality conformal fluid. In this regime, the dynamics is described by the constitutive relations truncated in the gradient expansion at the fixed order. The resulting forms of the stress energy tensor and (chiral) current read
\begin{align}
T_{\m\n}&=w u_\m u_\n + Pg_{\m\n}-\eta \s_{\m\n}+\sum_{i} \lambda_i\Sigma^{(i)}_{\m\n}\notag\\
J^\a&=n u^\a-\k P^{\a\b}\pa_{\b}\frac{\m}{T}+\xi_\o \o^\a+\sum_{i} \zeta_i\Theta^{(i)}_{\a}\,,
\end{align}
where $P^{\a\b}=g^{\a\b}+u^\a u^\b$, $\s^{\m\n}=P^{\m\a}P^{\n\b}(\pa_\a u_\b+\pa_\b u_\a)-\frac{2}{3}P^{\m\n}\pa^\a u_\a$, $\eta$ is the shear viscosity, $\kappa$ controls the charge diffusion, $\xi_\o$ is the chiral vortical conductivity, and the rest of the terms account for the contributions at the second order in hydrodynamic expansion.

Focusing on the linearized dynamics around a static background state, with $u^{(0)}=(1,0,0,0)$, constant temperature, and constant chemical potential, one notices that not all the second order tensorial structures are relevant. One may readily find that it is sufficient to consider five contributions that are linear in velocity, chemical potential, and pressure perturbations, namely
\begin{align}
&\Sigma^{(1)}_{\m\n}=P_{\m\a}P_{\n\b}u^\l\pa_\l\s^{\a\b}-\frac{1}{3}P_{\m\n}P^{\a\b}u^\l\pa_\l\s_{\a\b}\,,\notag\\ &\Sigma^{(2)}_{\m\n}=P_{\m\a}P_{\n\b}\pa^\a\pa^\b\frac{\m}{T}-\frac{1}{3}P_{\m\n}P^{\a\b}\pa_a\pa_\b\frac{\m}{T}\,, \notag\\
&\Sigma^{(3)}_{\m\n}=\frac{1}{2}P_{\m\a}P_{\n\b}(\pa^\a \o^\b+\pa^\b \o^\a)-\frac{1}{3}P_{\m\n}P^{\a\b}\pa_\a \o_\b\,,\notag\\
&\Theta^{(1)}_\a=P_{\a\b}\pa_\l\s^{\l\b}\,,~~~~~~\Theta^{(2)}_\a=P_{\a\b}\pa_\l\o^{\l\b}\,,
\end{align}
where $\o^{\m\n}=\frac{1}{2}P^{\m\a}P^{\n\b}(\pa_\a u_\b - \pa_\b u_\a)$, and for the explicit form of the background state we can further simplify $\Sigma^{(1)}_{\m\n}=\pa_t\s_{\m\n}$, $\Theta^{(1)}_\a=\pa^\l\s_{\l\a}$, and $\Theta^{(2)}_\a=\pa^\l\o_{\l\a}$. The only parity-odd structure among them is $\Sigma^{(3)}$, and the corresponding kinetic coefficient $\l_3$ is proportional to the single chirality anomalous coefficient $C_\chi$, where $\chi$ differentiates right-handed (R) and left-handed (L) particles. We will refer to this transport phenomenon as chiral vortical shear effect (CVSE), and this choice will become apparent later. Notice that when two chiralities are added, the corresponding contribution to the stress energy tensor becomes parity-even, with a parity-odd coefficient $\l_3\sim \m_5=\frac{1}{2}(\mu_R-\mu_L)$ since $C_R=-C_L$.

Projecting the linearized hydrodynamic equations with respect to the momentum, one finds that the longitudinal and transverse velocity components decouple. The anomaly affects only the latter ones, and the corresponding set of equations reads
\begin{align}
&\left(\o-i\frac{\eta}{w} k^2-\frac{\l_1}{w}\,\o\, k^2\right)\d\vec{v}_\perp\notag\\
&\hspace{2.5cm}+\frac{k^2}{4w}\l_3(\vec{k}\times\d\vec{v}_\perp)=0\,.
\label{treom}
\end{align}
Considering the determinant of this system, we find the dispersion relation of the collective modes in the transverse sector of the second-order chiral hydrodynamics: 
\begin{align}
\o=ik^2\frac{\eta \pm \frac{k}{4}\l_3}{w-\l_1 k^2}\,.
\label{eq:dispersions}
\end{align}
Remarkably, one of these modes is always unstable at sufficiently large momentum, indicating the tendency of the system to form spatially modulated flow configurations. 

These modes are known as CSWs \cite{Sahoo:2009yq}, but to the best of our knowledge, their role in flow helicity transfer and the associated unstable behavior has not been discussed before. In what follows, we focus on this chirality-driven instability of chiral media with respect to the generation of helical flows. To understand the properties of the above dispersion relations, it should also be noticed that the key coefficient $\lambda_3$ is constrained by the entropy-current analysis \cite{Kharzeev:2011ds}, as we discuss below.

The local velocity is not uniquely defined in the hydrodynamic expansion and can be freely adjusted with gradient corrections. Until this moment, we have been working in the Landau frame, fixing the particular form of the hydrodynamic description. Turning to the general case of the linearized second order chiral hydrodynamics, we have to consider a linear shift of the velocity leaving the charge and energy density unmodified
\begin{eqnarray}
u_\n\to u_\n +\a_1\, P_{\n\gamma}\pa^\gamma\frac{\mu}{T}  +\a_2\, \o_\n + \a_3\, \Theta^{(1)}_\n+\a_4\, \Theta^{(2)}_\n\,,\notag
\end{eqnarray}
where $\alpha$'s are free parameters that scale as corresponding powers of $w^{1/4}$. The modification of the spectrum of the transverse modes can be readily obtained, and we find
\begin{align}
\o=ik^2\frac{\eta\left(1\mp\frac{1}{2}\a_2k\right) \pm \frac{k}{4}\l_3}{w\left(1\mp\frac{1}{2}\a_2k\right)-\left(\l_1+\a_3w+\frac{1}{2}\a_4w\right) k^2}\,.
\end{align}
Thus, we see that the stability of the system is only slightly affected by such a frame shift as long as the momentum is sufficiently small. Indeed, expanding in powers of $k$ we can see that $\a_2$ cancels at the leading non-trivial order. Notice, however, that for the larger momentum modes the hydrodynamic description is not generally applicable, frame shifts are not allowed, and the theory should be specified microscopically.

\section{The properties of the Chiral Vortical Instability}

As we have seen in the previous section, CSWs become unstable in the Landau frame for large momentum, specifically when $k>4\eta/|\l_3|$, and provided that $w-\l_1 k^2>0$. This indicates that chiral media are generally unstable and tend to form spatially modulated topological flows, such as self-linked vortices. 

\begin{figure*}[t!]
    \centering
    \includegraphics[width=.497\textwidth]{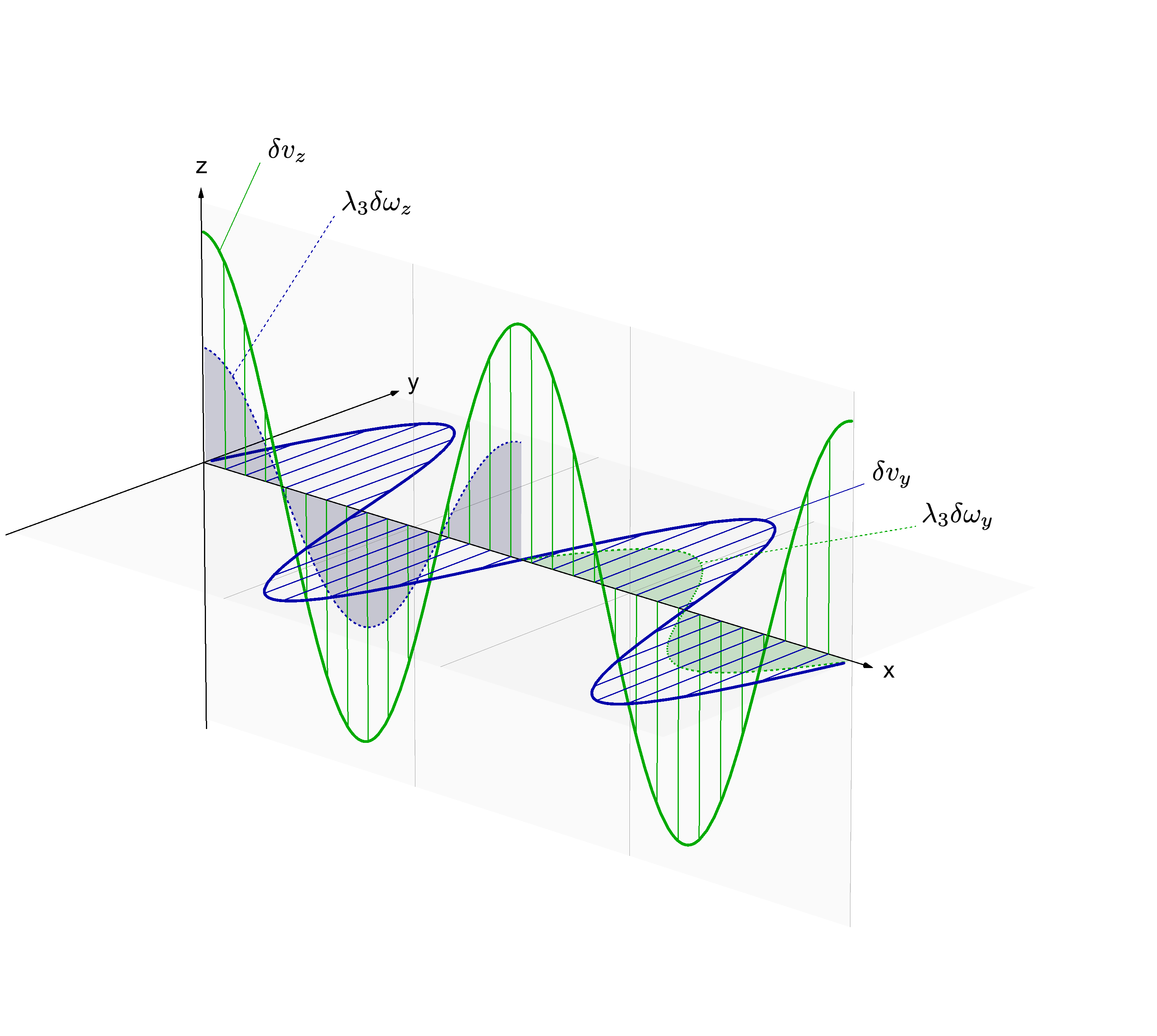}
    \includegraphics[width=.497
    \textwidth]{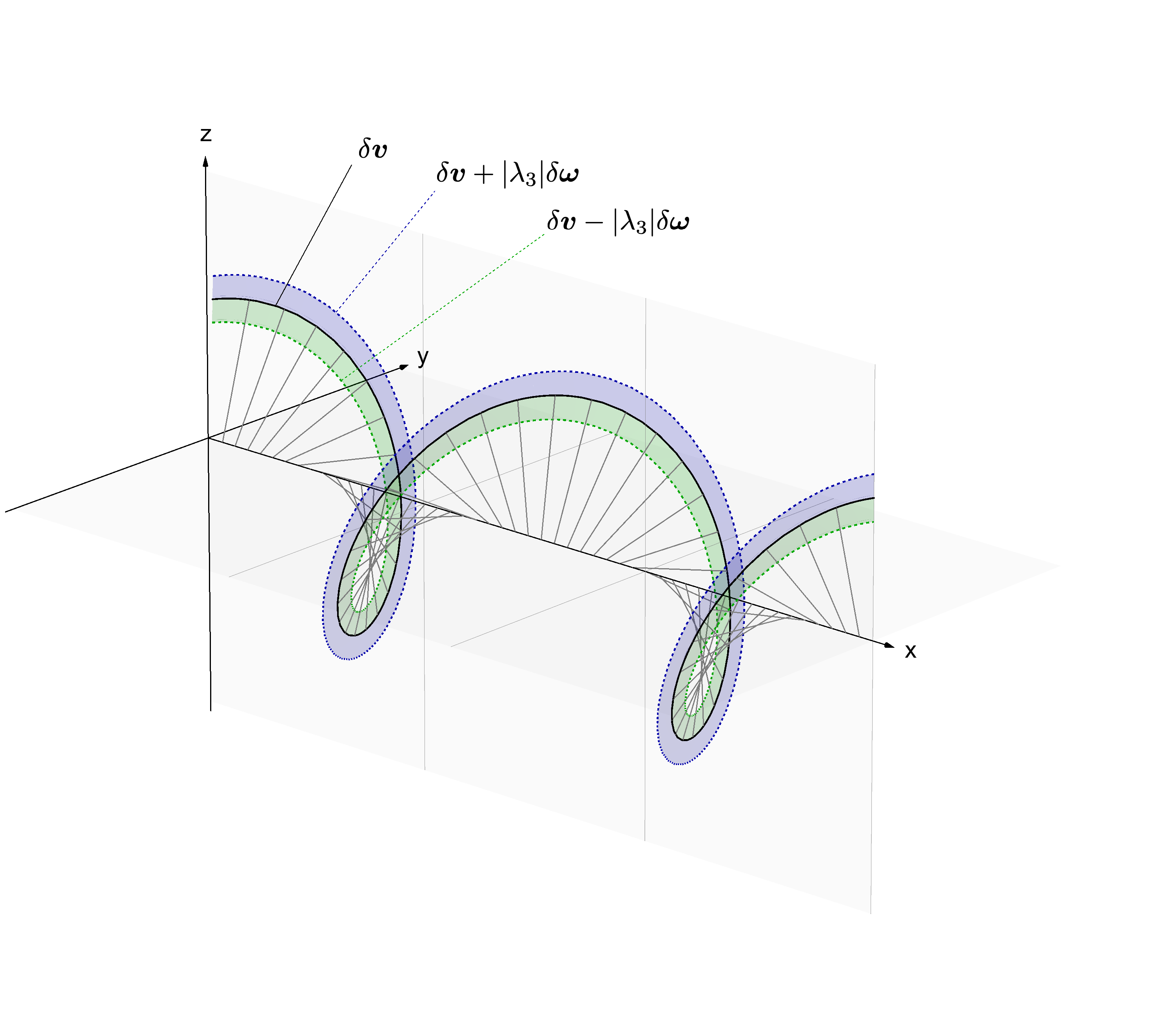}

    \caption{\textbf{Left:} This plot shows the velocity profile of the GABC solution in \eqref{GABC} at $t = 0$. Dashed curves show (scaled) vorticity contributions derived from the corresponding velocity components, color-coded accordingly. For better visualization, vorticity components are plotted in separate regions, assuming $\l_3>0$. The solid lines extending from the $x$-axis to the curves illustrate the velocity projections at a given $x$.
    \textbf{Right:} Velocity vectors with initial points on the $x$-axis follow a helical trajectory (solid line). The two colored bands illustrate how the flow velocity sums with the corresponding vorticity contributions in the effective shear term, depending on the sign of $\lambda_3$.
    }

    \label{fig1}
\end{figure*}

Indeed, solving for the zero vectors of \eqref{treom}, one can readily find that the elementary solutions are Gromeka–Arnold–Beltrami–Childress (GABC) flows \cite{Arnold2014, gromeka1881some,Beltrami1889,Childress1970}, for instance
\begin{equation}
\d \vec{v} = v_0 \left(
\begin{array}{c}
0\\
\sin k x_1\\
\cos k x_1
\end{array}
\right)e^{-t\,k^2\frac{\eta-\frac{k}{4}\l_3}{w-\l_1 k^2}}\,,
\label{GABC}
\end{equation}
where $v_0$ is an arbitrary amplitude, see Fig.~\ref{fig1} for an illustration. This flow satisfies $\delta {\bm \omega} = k \delta {\bm v}/2$, and $\Sigma_{\mu\nu}^{(3)}$ combines with $\sigma_{\m\n}$ in the stress energy tensor. The effective viscosity is given by $\eta - \lambda_3 k /4$, and the mode become unstable when it is negative. Such a flow has non-zero $\mathcal{H}_{fh}$, which grows exponentially in time. From the conservation of the generalized axial charge, we expect that a change in $\mathcal{H}_{fh}$ is compensated by a change in $n_5$, similar to the case of the CPI. Notice also that, on general grounds, a decline in $n_5$ leads to a decrease in $\m_5$, and, consequently, drives the system toward a stable state with a smaller $\lambda_3$ and non-zero $\mathcal{H}_{fh}$. 

At the limiting value of momentum, $k=4\eta/|\l_3|$, the linearized hydrodynamic description of chiral media supports static topological flows. Strikingly, such solutions of linearized second order chiral hydrodynamics can, in fact, be generalized to the fully non-linear regime. Indeed, a direct substitution shows that a {\it static} GABC flow satisfies the hydrodynamic equations for
\begin{eqnarray}
\label{CVEnl}
k=\frac{4\eta}{|\l_3|}\sqrt{1-v_0^2}
\,.
\end{eqnarray}
The form of this solution suggests that nonlinear effects, at least up to second order in gradients, do not suppress the instability. Moreover, it is important to emphasize that global rotation is not a solution of second-order chiral hydrodynamics. Instead, it is expected to deform into a static GABC solution with a fixed momentum \eqref{CVEnl}, which satisfies the equilibrium constraint--provided that $\m_5$ is allowed in equilibrium and in the absence of electromagnetic fields, see e.g.~\cite{Becattini:2012tc, Becattini:2013fla}. Such a modification of the general equilibrium state requires further study, which we leave for future work.

We note in passing that CPI and CVI generate long- and short-wavelength excitations, respectively, and that helical field and flow configurations are typically generated at opposite length scales. It would be interesting to study not only the cascade of CVI alone but also the interplay among the three instabilities.

Let us now briefly comment on the validity of the hydrodynamic expansion. While large-momentum modes cannot be reliably described within the hydrodynamic framework, this is generally the case for hydrodynamic vortices, which are widely discussed in the literature and observed experimentally. However, they are usually treated either through simulations at a fixed order or in general terms of symmetries and conservation laws, see e.g. \cite{Endlich:2010hf,Endlich:2013dma,Montenegro:2017rbu,Cuomo:2024ekf}. Here, we argue that the modification of the equilibrium state of chiral matter—favoring helical flow configurations while excluding global rotation—strongly suggests the presence of CVI beyond the applicability of the hydrodynamic description utilized in this work.

In turn, if $\eta/|\lambda_3|$ becomes sufficiently small, this would indicate that CVI also occurs for hydrodynamic long-wavelength modes that are not strongly affected by higher-order gradient corrections. However, the corresponding upper bound depends on how the hydrodynamic description is treated in the absence of microscopic picture. For instance, one may constrain the momentum by comparing the zeroth-order stress-energy tensor with the first few gradient corrections, ensuring that $w\gg \text{max}\left(\eta\,k, |\lambda_3|\,k^2\right)$, or by treating it as a characteristic scale—excluding the particular kinetic coefficients—by requiring $k\ll w^{1/4}$. Notice that in principle one may also involve the amplitude of the linearized mode in such a comparison, but we skip that option here, avoiding further issues with the higher order non-linear terms mixing the gradient and amplitude expansions.

To further constrain the limiting value of the momentum, we have to notice that some of the anomalous coefficients can be constrained solely from the second law of thermodynamics \cite{Son:2009tf,Sadofyev:2010pr,Neiman:2010zi}. Following \cite{Kharzeev:2011ds}, we find that
\begin{eqnarray}
&\lambda_3=-\frac{4\eta}{w} \left( \frac{1}{3}C_{\chi} \m^3  + \beta_\chi  \m T^2  \right)\,,
\end{eqnarray}
where $C_\chi$ and $\beta_\chi$ are the coefficients of the regular and mixed gauge-gravitational axial anomalies, respectively,
whose specific values depend on (chiral) underlying microscopic theory \cite{Landsteiner:2011cp}. It should be stressed here that since $\l_3$ scales with $\eta$ it differs from the first-order anomalous kinetic coefficients. Indeed, the leading chiral anomalous effects such as CVE are field-theoretical phenomena, arising at the microscopic level, and are, thus, non-dissipative. In turn, CVSE has no simple microscopic counterpart, and the particular form of $\l_3$ constrained by the hydrodynamic considerations indicates that it appears as a result of fixing the hydrodynamic frame, see e.g. \cite{Landsteiner:2012kd,Rajagopal:2015roa,Stephanov:2015roa}. Inserting the explicit form of $\l_3$ into the stability condition, we find that the characteristic momentum is $\eta$-independent and reads
\begin{eqnarray}
k>\left|\frac{3w}{C_\chi \m^3+3\beta_\chi \m T^2}\right|\,.
\end{eqnarray}

To illustrate how CVI can be realized in the hydrodynamic regime, one can consider a free dense fermion gas with $w=\frac{N_f}{6\pi^2}\m^4$ and $C_\chi=\frac{N_f}{4\pi^2}$, where $N_f$ is the number of fermion flavors. Then, if we require that $k\ll w^{1/4}$ and $N_f\gg1$, the hydrodynamic modes with
\begin{eqnarray}
2\sqrt[4]{6\pi^2/N_f}<k/w^{1/4}\ll 1\notag
\end{eqnarray}
become unstable. Increasing $N_f$, we increase the anomalous coefficient, and the CVI can be identified with the helical instability of the holographic plasma, which transitions into a spatially modulated phase for a large Chern-Simons coupling of the dual description controlling the anomaly \cite{Nakamura:2009tf}. Similarly to the case of large $N_f$, one may consider other setups supporting non-trivial anomalous coefficients, e.g. focusing on non-Abelian media.

If instead we directly compare the first two gradient corrections with the background thermodynamic scale, requiring $w> \text{max}\left(\eta\,k, |\lambda_3|\,k^2\right)$, we readily find that CSWs could become unstable only for $\eta\,k<|\lambda_3|\,k^2$, as long as the momentum satisfies
\begin{eqnarray}
\frac{4\eta}{|\lambda_3|}<k<\sqrt{\frac{w}{|\lambda_3|}}\,.\notag
\end{eqnarray}
This range is non-trivial if $4\eta<\frac{1}{3}C_\chi\m^3+\b_\chi\m T^2$, and one may further conclude that nearly ideal chiral fluids generally tend to exhibit CVI even for sufficiently long wavelength modes.

\section{Conclusions and outlook}

We have revisited the spectrum of linearized second-order chiral hydrodynamics and identified a chiral instability associated with CSWs for $|k|>|4\eta/\l_3|$, leading the system to support topological helical flows. Its explicit form reveals a mechanism that transforms microscopic chirality $N_5$ into macroscopic helicity $\mathcal{H}_{fh}$, which we call the {\it chiral vortical instability}. The elementary flow induced by the CVI takes the form of GABC solutions with exponentially growing amplitudes. Thus, the CVI, similar to the CPI and CMVI, can be thought of as following from the necessity of redistributing generalized axial charge when the dynamics is initiated at finite $\mu_5$ but with zero electromagnetic field and flow \cite{Avdoshkin:2014gpa}.

The unstable modes grow faster with larger momenta, indicating that chiral media tend to generate a spatially modulated flow, e.g. a phase with linked vortices. Such solutions can only be fully described within a microscopic framework, completing the hydrodynamic picture at short distances. For instance, it would be interesting to explore the large-momentum behavior of the unstable modes in an all-order hydrodynamic framework \cite{Bu:2014sia, Bu:2014ena} or within chiral kinetic theory, see \cite{Hidaka:2022dmn} for a recent review. Alternatively, one may focus on chiral superfluidity and the realization of chiral anomalous effects on microscopically controlled superfluid vortices, see e.g. \cite{Kirilin:2012mw, Hirono:2018bwo}. Here, we only argue that since the static GABC flow solves the full nonlinear second-order hydrodynamics, one may expect that CVI is not immediately regulated by the non-linear effects. This also highlights that global rotation is not an equilibrium state of chiral media as long as $\mu_5\neq0$. We show further that CSWs can become unstable even for momenta within the hydrodynamic regime. Lacking microscopic control over the theory, we explore two potential realizations of this scenario with different upper momentum cutoffs.

The conservation of the generalized axial charge provides insight into the dynamics of chiral media. However, since the anomaly remains unmodified in the presence of matter, one may wonder how chirality could transform into a helical flow. A potential microscopic realization of such a transfer may involve the polarization of gauge fields in rotating matter, see e.g. \cite{Avkhadiev:2017fxj, Yamamoto:2017uul, Huang:2018aly, Chernodub:2018era, Huang:2020kik, Hattori:2020gqh}. The details of this transfer in the case of CVI are left for future studies within a suitable microscopic completion.

Our findings suggest a variety of possible phenomenological implications in chiral fluids. First, these systems will naturally develop spontaneous topological flows and linked vortex configurations due to the presence of CVI in their dynamics. In the context of the quark-gluon plasma (QGP) produced in heavy-ion collisions, CVI may influence the final-state local spin polarization \cite{Gao:2020lxh, Huang:2020dtn, Becattini:2024uha}, which is sensitive to the QGP vortices, see e.g. \cite{Lisa:2021zkj, DobrigkeitChinellato:2024xph}. In the primordial plasma of the early Universe, CVI may affect the generation of magnetic fields and the global rotation of a variety of structures \cite{Kamada:2022nyt}, and may lead to observable imprints in the modern Universe, e.g. in rotation of galaxies \cite{Yu:2019bsd}. In the interior of neutron stars and supernovae, it could influence neutrino transport \cite{Yamamoto:2015gzz} and contribute to pulsar kicks \cite{Shaverin:2014xya,Kaminski:2014jda}. Second, wherever chiral matter tends to rotate, CVI will destabilize the system, leading to the formation of helical flow configurations and catalyzing the transition to turbulence. 

Finally, it should be mentioned that chiral hydrodynamics has been argued to be unstable and/or develop features violating causality \cite{Speranza:2021bxf,Abboud:2023hos} at the level of its equations of motion, and it would be interesting to relate the variety of chiral instabilities with such considerations.

\section*{Acknowledgments}
We thank E. Speranza, G. Torrieri, and P. Zakharevsky for useful comments. The work of A.V.S is supported by Funda\symbol{"00E7}\symbol{"00E3}o para a Ci\symbol{"00EA}ncia e a Tecnologia (FCT) under contracts 
2023.15319.PEX (https://doi.org/10.54499/2023.15319.PEX) and 2022.06565.CEECIND. The works of S.W. and X.G.H are supported by Natural Science Foundation of Shanghai through Grant No. 23JC1400200, National Natural Science Foundation of China through Grants No. 12225502, No. 12147101, and National Key Research and Development Program of China through Grant No. 2022YFA1604900. 
The work of K.H. is partially supported by the JSPS KAKENHI under grant No. 22H01216.

\bibliographystyle{bibstyle}
\bibliography{references}

\end{document}